\documentclass[prd,email,twocolumn,aps,showpacs,showkeys,preprintnumbers,amsmath,amssymb,nofootinbib]{revtex4-1}

\def\[{\left\lbrack}
\def\]{\right\rbrack}

\def\({\left(}
\def\){\right)}

\newcommand{\be}{\begin{equation}}
\newcommand{\ee}{\end{equation}}
\newcommand{\ea}{\end{eqnarray}}
\newcommand{\ba}{\begin{eqnarray}}

\begin{document}

\title{Lagrangian formulation for noncommutative nonlinear systems}

\author{E. M. C. Abreu$^{a,b}$}
\email{evertonabreu@ufrrj.br}
\author{J. Ananias Neto$^{c}$}
\email{jorge@fisica.ufjf.br}
\author{A.C.R. Mendes$^{c}$}
\email{albert@fisica.ufjf.br} 
\author{C. Neves$^{d}$}
\email{cliffordneves@uerj.br}
\author{W. Oliveira$^{c}$}
\email{wilson@fisica.ufjf.br}
\author{M. V. Marcial}
\email{mateusmarcial@fisica.ufjf.br}

\affiliation{$^a$Grupo de F\' isica Te\'orica, Departamento de F\'{\i}sica, \\
Universidade Federal Rural do Rio de Janeiro\\
BR 465-07, 23890-971, Serop\'edica, RJ, Brazil\\
$^b$Centro Brasileiro de Pesquisas F\' isicas (CBPF), Rua Xavier Sigaud 150,\\
Urca, 22290-180, RJ, Brazil\\
$^c$Departamento de F\'{\i}sica, ICE, Universidade Federal de
Juiz de Fora, 36036-330, Juiz de Fora, MG, Brasil\\
$^d$Departamento de Matem\'atica, F\'\i sica e Computa\c{c}\~ao, Universidade do Estado do Rio de Janeiro,\\
Rodovia Presidente Dutra, km 298, 27537-000, Resende, Rio de Janeiro, Brazil\\\\
\today\\}


\begin{abstract}
\noindent
In this work we use the well known formalism developed by Faddeev and Jackiw to introduce noncommutativity within two nonlinear systems, the SU(2) Skyrme and O(3) nonlinear sigma models.  The final result is the Lagrangian formulations for the noncommutative versions of both models.  The possibility of obtaining different noncommutative versions for these nonlinear systems is demonstrated.
\end{abstract} 
 
\maketitle

\pagestyle{myheadings}
\markright{\it Lagrangian formulation for noncommutative nonlinear systems}

\section{Introduction}

In current theoretical physics there is a relevant number of theoretical investigations that lead us to believe that at the first moments of the Big-Bang, the geometry was not commutative and the dominating physics at that time was ruled by the laws of noncommutative (NC) geometry.  Therefore, the idea that the physics of the early moments can be constructed based on these concepts.  

The first published steps through this knowledge were given by Snyder \cite{snyder} which believes that NC principles could make the quantum field theory divergences disappear.  However, it was not accomplished \cite{yang} and Snyder's ideas were put to sleep for a long time.

The main modern motivations that rekindle the investigation about  NC field theories  come from string theory and quantum gravity \cite{strings}.   

In the context of quantum mechanics for example, R. Banerjee \cite{banerjee} discussed how noncommutative structures appear in planar quantum mechanics providing a useful way for obtaining them.   The analysis was based on 
the noncommutative algebra in planar quantum mechanics that was originated from 't Hooft's analysis on dissipation and quantization \cite{thooft}. 

It is opportune to mention here that this noncommutativity in the context of string theory mentioned above could be eliminated constructing a mechanical system which reproduces the string classical dynamics \cite{string}.   NC field theories have been studied intensively in many branches of physics \cite{witten,other,alexei,belov,szabo,omer,RB,hp}.

In a very interesting paper, a parallel investigation was developed by Duval and Horv\'athy \cite{dh}, where it was obtained the anomalous commutation relations for the coordinates obtained through the ``Peierls substitution" \cite{peierls}.  From first principles, without using such unphysical limit, the authors introduced NC (quantum) mechanics starting with group theory and applied it to condensed matter physics, e.g., the Hall effect.  The respective Lagrangian approach was discussed in detail in subsequent papers \cite{dh2}.  Dunne, Jackiw and Trugenberger \cite{djt} justify the Peierls rule by considering the $m\rightarrow 0$ limit, reducing classical phase space from four to two dimensions, parameterized by NC coordinates $X$ and $Y$, whereas the potential becomes an effective Hamiltonian.

The perturbative study of scalar field theories was performed in \cite{mrs}.  The authors analyzed the IR and UV divergences and verified that Planck's constant enters via loop expansion.  Here, differently, we make a non-perturbative approach and we will see that Planck's constant enters naturally in the theory via Moyal-Weyl product.

In \cite{DJEMAI1}, a general algebra $\alpha$-deformation of classical observables that introduces a general NC quantum mechanics was constructed.  This $\alpha$-deformation is equivalent to some general transformation for the usual quantum phase space variables.  In other words, the authors discuss the passage from classical mechanics to quantum mechanics.  Then to NC quantum mechanics, which allows to obtain the associated NC classical mechanics. This is possible since quantum mechanics is naturally interpreted as a NC (matrix) symplectic geometry \cite{DJEMAI2}.

In \cite{amorim}, the author constructed an extension of the well known Doplicher-Fredenhagen-Roberts NC algebra introducing the formalism which is now called in the literature as the Doplicher-Fredenhagen-Roberts-Amorim algebra.  In this formalism the NC parameter $(\theta)$ is an ordinary coordinate of the spacetime and therefore it has a canonical conjugate momentum $(\pi)$.  An extended Hilbert space was constructed together with all the ingredients of a new NC quantum mechanics.  This formalism is an alternative to the Moyal-Weyl one.  But notice that both preserves the underlying NC relation $[x^\mu ,x^\nu ]=i\theta^{\mu\nu}$.  For details, the interested reader can consult \cite{sigma}.

Back to our main subject here, in few words we can say that to obtain NC versions for field theories one have to replace the usual product of fields into the action by the Moyal-Weyl product, defined as
\be
\phi_1 (x)\star \phi_2 (x) \,=\,exp\left( {i\over 2} \theta^{\mu \nu}
\partial^{x}_{\mu}\partial^{y}_{\nu} \right) \phi_1 (x) \phi_2 (y)\mid_{x=y}, 
\nonumber
\ee
where $\theta^{\mu \nu}$ is a real and antisymmetric constant matrix. As a consequence, NC theories are highly nonlocal. We also note a basic NC property that the Moyal-Weyl product of two fields inside the action is the same as the usual product, considering that we discard boundary terms. Thus, the noncommutativity affects just the vertices.

In \cite{sympnc} three of us have proposed a new formalism to generalize the quantization by deformation introduced in \cite{DJEMAI1} in order to explore, with a new insight, how the NC geometry can be introduced into a (commutative) field theory. To accomplish this, a systematic way to introduce NC geometry into commutative systems, based on Faddeev-Jackiw symplectic formalism and Moyal product, was presented \cite{amo}. 

Further, this method describes precisely how to obtain a Lagrangian description for the NC version of the system. In this work we apply this formalism to obtain Lagrangian formulations for NC versions for $SU(2)$ Skyrme model and $O(3)$ nonlinear sigma model. 

We have organized this paper as follows. In Section II, we introduce our generalized quantization by deformation assuming a generic classical symplectic structure.  In section III we show how to construct Lagrangian formulations for NC versions for the following nonlinear systems: the $SU(2)$ Skyrme Model and $O(3)$ sigma model.  The conclusion is depicted in last section.

\section{The NC Generalized Symplectic Formalism}

The quantization by deformation \cite{MOYAL1} consists in the substitution of the canonical quantization process by the algebra ${\cal A}_\hbar$ of quantum observables generated by the same classical one obeying Moyal-Weyl product, {\it i.e.}, the canonical quantization
\be
\label{II1}
\lbrace h, g\rbrace_{PB} = \frac{\partial h}{\partial \zeta_a} \omega_{ab} \frac{\partial g}{\partial \zeta_b}  \longrightarrow \frac {1}{\imath\hbar} [{\cal O}_h, {\cal O}_g]\;\;,
\ee

\noindent with $\zeta=(q_i,p_i)$, is replaced by the $\hbar$-star deformation of ${\cal A}_0$, given by

\be
\label{II2}
\lbrace h, g\rbrace_{\hbar} = h*_\hbar g - g*_\hbar h\;\;,
\ee
where
\be
\label{II3}
(h *_\hbar g)(\zeta)\,=\,\exp\{\frac{\imath}{2}\hbar \omega_{ab}\partial^a_{(\zeta_1)}\partial^b_{(\zeta_2)}\}h(\zeta_1)g(\zeta_2)|_{\zeta_1=\zeta_2=\zeta}
\;\;,
\ee
with $a,b=1,2,\dots,2N$ and with the following classical symplectic structure
\be
\label{II4}
\omega_{ab} = \left( \begin{array}{cc}
0 & \delta_{ij} \\ 
-\delta_{ji} & 0
\end{array} \right)
\,\,\,{\text with}\,\,\, i,j=1,2,\dots,N\;\;,
\ee
which satisfies the relation
\be
\label{II5}
\omega^{ab}\omega_{bc} = \delta^a_c\;\;.
\ee

The quantization by deformation can be generalized assuming a generic classical symplectic structure $\Sigma^{ab}$. In this way the internal law will be characterized by $\hbar$ and by another deformation parameter (or more). As a consequence, the $\Sigma$-star deformation of the algebra becomes
\be
\label{II6}
(h *_{\hbar\Sigma} g)(\zeta)\,=\, \exp\{\frac{\imath}{2}\hbar \Sigma_{ab}\partial^a_{(\zeta_1)}\partial^b_{(\zeta_2)}\}h(\zeta_1)g(\zeta_2)|_{\zeta_1=\zeta_2=\zeta}\,\,,
\ee
with $a,b=1,2,\dots,2N$.

This new star-product generalizes the algebra among the symplectic variables  in the following way
\be
\label{II7}
\lbrace h, g\rbrace_{\hbar\Sigma} = \imath\hbar\Sigma_{ab}\;\;.
\ee

In \cite{DJEMAI1,DJEMAI2}, the authors proposed a quantization process to transform the NC classical mechanics into the NC quantum mechanics, through generalized Dirac quantization,
\be
\label{II8}
\lbrace h, g\rbrace_{\Sigma} = \frac{\partial h}{\partial \zeta_a} \Sigma_{ab} \frac{\partial g}{\partial \zeta_b}  \longrightarrow \frac {1}{\imath\hbar} [{\cal O}_h, {\cal O}_g]_{\Sigma}\;\;.
\ee

\noindent The relation above can also be obtained through a particular transformation onto the usual classical phase space, namely,
\be
\label{II9}
\zeta^\prime_a = T_{ab} \zeta^b\;\;,
\ee
where the transformation matrix is
\be
\label{II10}
T = \left( \begin{array}{cc}
 \delta_{ij} & - \frac 12 \theta_{ij} \\  
\frac 12 \beta_{ij}  & \delta_{ij}
\end{array} \right)\;\;,
\ee
with $\theta_{ij}$ and $\beta_{ij}$ being antisymmetric matrices. As a consequence, the original Hamiltonian becomes
\be
\label{II11}
{\cal H}(\zeta_a) \longrightarrow {\cal H}(\zeta^\prime_a)\;\;.
\ee
The corresponding symplectic structure is
\be
\label{II12}
\Sigma_{ab} = \left( \begin{array}{cc}
\theta_{ij} & \delta_{ij}+\sigma_{ij} \\ 
-\delta_{ij}-\sigma_{ij} & \beta_{ij}
\end{array} \right) \;\;,
\ee
$\sigma_{ij} = - \frac18 [\theta_{ik}\beta_{kj} + \beta_{ik}\theta_{kj}]$. Due to this, the commutator relations look like
\ba
\label{II13}
\[q^\prime_i, q^\prime_j\] &=& \imath\hbar\theta_{ij}\;\;,\nonumber\\
\[q^\prime_i, p^\prime_j\] &=& \imath\hbar (\delta_{ij} + \sigma_{ij})\;\;,\\
\[p^\prime_i, p^\prime_j\] &=& \imath\hbar\beta_{ij}\;\;.\nonumber
\ea

At this point, it is important to notice that a Lagrangian formulation was not given so far. Now, we propose a new systematic way to obtain a NC Lagrangian description for a commutative system. In order to achieve our objective, the symplectic structure $\Sigma_{ab}$ must firstly be fixed and after that, the inverse of $\Sigma_{ab}$ must be computed. As a consequence, an interesting problem arise: if there are some constants ({\it Casimir invariants}) in the system, the symplectic structure has a zero-mode, given by the gradient of these {\it Casimir invariants}. Hence, it is not possible to compute the inverse of $\Sigma_{ab}$. However, in Ref. \cite{CNWO} this kind of problem was solved. On the other hand, if $\Sigma_{ab}$ is nonsingular, its inverse can be obtained solving the next relation
\be
\label{II14}
\int{\Sigma_{ab}(x,y) \, \Sigma^{bc}(y,z) dy}\, = \,\delta_a^c \delta(x-z)\;\;,
\ee

\noindent which generates a set of differential equations since $\Sigma^{ab}$ is an unknown two-form symplectic tensor obtained from the following first-order Lagrangian
\be
\label{II15}
{\cal L} = A_{\zeta^\prime_a} \dot\zeta^{\prime a} - V(\zeta^\prime_a)\;\;,
\ee
as being
\be
\label{II16}
\Sigma^{ab}(x,y) = \frac {\delta A_{\zeta^\prime_a}(x)}{\delta \zeta^\prime_b(y)} - \frac {\delta A_{\zeta^\prime_b}(x)}{\delta \zeta^\prime_a(y)}\;\;.
\ee

\noindent Due to this, the one-form symplectic tensor, $A_{\zeta^\prime_a}(x)$, can be computed and subsequently, the Lagrangian description, Eq. (\ref{II15}), is obtained also. 

In order to compute $A_{\zeta^\prime_a}(x)$, the Eq. (\ref{II14}) and Eq. (\ref{II16}) will be used, which generates the following set of differential equations
\ba
\label{II17}
\theta_{ij} B_{jk}(x,y) + \(\delta_{ij}+\sigma_{ij}\)A_{jk}(x,y) &=& \delta_{ik}\delta(x-y)\;\;,\nonumber\\
A_{jk}(x,y) \theta_{ji} + \(\delta_{ij}+\sigma_{ij}\)C_{jk}(x,y) &=& 0\;\;,\nonumber\\
- \(\delta_{ij} + \sigma_{ij}\)B_{jk}(x,y) + \beta_{ij}A_{jk}(x,y) &=& 0\;\;, \\
A_{kj}(x,y)\(\delta_{ji} + \sigma_{ji}\) + \beta_{ij} C_{jk}(x,y) &=& \delta_{ik}\delta(x-y)\;\;,\nonumber 
\ea
where
\ba
\label{II18}
B_{jk}(x,y) &=& \(\frac {\delta A_{q^\prime_j}(x)}{\delta q^\prime_k(y)} - \frac {\delta A_{q^\prime_k}(x)}{\delta q^\prime_j(y)}\)\;\;,\nonumber\\
A_{jk}(x,y) &=& \(\frac {\delta A_{p^\prime_j}(x)}{\delta q^\prime_k(y)} - \frac {\delta A_{q^\prime_k}(x)}{\delta p^\prime_j(y)}\)\;\;,\nonumber\\
C_{jk}(x,y) &=& \(\frac {\delta A_{p^\prime_j}(x)}{\delta p^\prime_k(y)} - \frac {\delta A_{p^\prime_k}(x)}{\delta p^\prime_j(y)}\)\;\;.
\ea

\noindent From the set of differential equations in Eq. (\ref{II17}), and the equations above, Eq. (\ref{II18}), we compute the quantities $A_{\zeta^\prime_a}(x)$.

As a consequence, the first-order Lagrangian can be written as

\be
\label{II19}
{\cal L} = A_{\zeta^\prime_a} \dot\zeta^\prime_a - V(\zeta^\prime_a)\;\;.
\ee
Notice that, despite (\ref{II15}) and (\ref{II19}) have the same form, in (\ref{II19}) the $A_{{\zeta'}_a}$ are completely computed through the solution of the system (\ref{II17}).   In both we have a NC version of the theory as a consequence of the deformation in (\ref{II10}) and its corresponding symplectic structure in (\ref{II12}).

\section{Lagrangian formulation for noncommutative nonlinear systems}

\subsection{The Noncommutative $SU(2)$ Skyrme Model}

The Skyrme model describes baryons and their interactions through soliton solutions of the nonlinear sigma model. The classical static Lagrangian for the Skyrme model is given by

\ba
\label{sk1}
L &=& \int d^{3}x \{- \frac{f_{\pi}^{2}}{16} Tr(\partial_\mu U \partial^\mu U^{\dag}) \nonumber\\
&+& \frac{1}{32e^2} Tr \[U^{\dag}\partial_\mu U, U^{\dag}\partial_\nu U \]^2 \},
\ea

\noindent where $f_{\pi}$ is the pion decay constant, $e$ is a dimensionless parameter and $U$ is an $SU(2)$ matrix. Performing the collective semi-classical expansion \cite{sky1}, substituting $U(r)$ by $U(r,t)=A(t)U(r)A^{\dag}(t)$ in (\ref{sk1}), where $A$ is an $SU(2)$ matrix, we obtain

\be
\label{sk2}
L = -M + \lambda Tr\[\partial_{0}A\partial_{0}A^{-1}\],
\ee

\noindent where $M$ is the soliton mass and $\lambda$ is the moment of inertia \cite{sky1}. The $SU(2)$ matrix $A$ can be written as $A = a_{0} + i a\cdot\tau$, where $\tau_{i}$ are the Pauli matrices, and satisfies the spherical constraint relation $T_{1} = a_{i}a_{i} - 1 \approx 0,\;\;\; i = 0,1,2,3.$



Performing the canonical quantization by using the Faddeev-Jackiw formalism \cite{sky2} we obtain the twice-iterated Lagrangian

\be
\label{sk3}
L = \(\pi_{i} + \rho a_{i} + \eta \pi_{i} \)\dot{a_{i}} + \eta a_{i} \dot{\pi_{i}} - V,
\ee
\noindent where

\be
\label{sk4}
V = M + \frac{1}{8\lambda} \pi_{i}\pi_{i},
\ee

\noindent and $\rho$ and $\eta$ are Lagrangian multipliers.

The inverse of the symplectic matrix that gives the usual Dirac brackets of the physical variables is given by \cite{sky2}

\be
\label{sk5}
\Sigma^{-1}=\left(\begin{array}{llll}
\qquad0 & \:\:\delta_{ij} - a_{i}a_{j} & \:\:a_{i} &\:\: 0\\
-\delta_{ji} + a_{j}a_{i}&\:\: a_{j}\pi_{i} - a_{i}\pi_{j}& \:\:-\pi_{i} &\:\:\: a_{i}\\
\qquad- a_{j} & \qquad\pi_{j} & \:\:0 &\:\: -\delta_{ij}\\
\qquad 0 & \qquad-a_{j} &\:\: \delta_{ji} &\quad 0
\end{array}\right). 
\ee

The Dirac brackets are then given by

\ba
\label{sk6}
\lbrace a_{i},a_{j}\rbrace &=& 0,\nonumber\\
\lbrace a_{i},\pi_{j}\rbrace &=& \delta_{ij} - a_{i}a_{j},\\
\lbrace \pi_{i},\pi_{j}\rbrace &=& a_{j}\pi_{i} - a_{i}\pi_{j}.\nonumber\\
\ea

In order to disclose the noncommutative version of the $SU(2)$ Skyrme model, we start proposing a new bracket relation between the collective coordinates that in symplectic language we have

\begin{widetext}
\be
\label{sk7}
\Sigma^{-1}=\left(\begin{array}{llll}
\qquad\theta_{ij} &\:\: \delta_{ij} - a_{i}a_{j} &\:\: a_{i} &\:\: 0\\
-\delta_{ji} + a_{j}a_{i} &\:\: a_{j}\pi_{i} - a_{i}\pi_{j}&\:\: -\pi_{i} &\:\: a_{i}\\
\qquad- a_{j} &\qquad \pi_{j} &\:\: 0 & \:\:-\delta_{ij}\\
\qquad 0 & \qquad -a_{j} & \:\:\delta_{ji} &\:\: 0
\end{array}\right), 
\ee

\noindent where $\theta_{ij}$ is an antisymmetric matrix. After a straightforward computation, the symplectic matrix is given by

\be
\label{sk8}
\Sigma = \left(\begin{array}{llll}
0 & \qquad-\delta_{ij}  & \qquad\qquad-a_{i} & \qquad\qquad-\pi_{i}\\
\delta_{ji} & \qquad\theta_{ij} & \qquad\qquad\theta_{ij}a_{j} & \qquad\quad\theta_{ik}\pi_{k}-a_{i}\\
a_{j} & \qquad a_{k}\theta_{kj} & \qquad\qquad\quad 0 & \:\:1-a_{k}a_{k}-\pi_{l}\theta_{lk}a_{k}\\
\pi_{j} & \:\:a_{j} - \pi_{l}\theta_{lj}&\:\: -1 + a_{k}a_{k} + \pi_{l}\theta_{lk}a_{k} &\qquad\qquad 0
\end{array}\right). 
\ee
\end{widetext}

\noindent From the definition of symplectic matrix, 

\be
\label{sk9}
\Sigma_{\xi_{i}\xi_{j}} = \frac{\partial A_{\xi_{i}}}{\partial \xi_{j}} - \frac{\partial A_{\xi_{j}}}{\partial \xi_{i}},
\ee

\noindent we obtain a set of differential equations that allows us to compute all the one-forms canonical momenta $A_{\xi_{i}}$, Eqs. (\ref{II17}) and  Eq. (\ref{II18}). Thus, we have

\ba
\label{sk10}
A_{a_{i}} &=& \pi_{i} +\rho a_{i} + \eta \pi_{i} - \rho \partial_{a_{i}} \(\theta_{kl}a_{l}\pi_{k} \), \nonumber \\
A_{\pi_{i}} &=& \eta a_{i} + \frac{1}{2}\theta_{ji}\pi_{j} - \rho \theta_{ij} a_{j}, \\
A_{\rho} &=& 0, \nonumber \\
A_{\eta} &=& 0.
\ea

\noindent These solutions were obtained considering the following relations $\theta_{ij} \pi_{j} \approx 0$ and $1 - a_{i}a_{i} - a_{k} \pi_{l} \theta_{lk} \approx 0$.  

Then the first-order Lagrangian which describe the NC $SU(2)$ Skyrme Model is

\begin{widetext}
\be
\label{sk11}
L = \( \pi_{i} +\rho a_{i} + \eta \pi_{i} - \rho \partial_{a_{i}}( \theta_{kl}a_{l}\pi_{k}) \)\dot{a_{i}} + ( \eta a_{i} + \frac{1}{2}\theta_{ji}\pi_{j} - \rho \theta_{ij} a_{j} )\dot{\pi_{i}} - V.
\ee   
\end{widetext}

Analyzing (\ref{sk11}) we can write 
\be
\label{sk10023}
L = \( \pi_{i} +\rho \tilde{a}_{i} + \eta \pi_{i} \)\dot{a_{i}} + ( \eta {a'}_i )\dot{\pi_{i}} - V\,\,,
\ee
where 
\ba
\tilde{a}_i \,&=&\, a_i\,-\,\partial_{a_i} (\theta_{kl}\,a_l \pi_k) \\
\eta{a'}_i\,&=&\,\eta\,a_i\,-\,\frac 12\, \theta_{ij}\,\bar{a}_j \\
\bar{a}_i\,&=&\,\pi_j \,-\,2 \rho\,a_j\,\,, 
\ea
and the action (\ref{sk10023}) is analogous to (22).

However, comparing both (22) and (32) it is easy to see that the noncommutativity modifies directly the elements $a_i$ of the $SU(2)$ matrix $A$, as seen in (34) and (35).  During the introduction of the Faddeev-Jackiw formalism, new symplectic variables were introduced, but the noncommutativity act only in $a_i$ terms in both $\dot{a}_i$ and $\dot{\pi}$ sectors.  The NC parameter also ``couples" with the $\pi$ factors. 
But there is no NC contribution in the potential sector, where the noncommutativity does not modify the soliton mass, which would be an interesting result.  Moreover, we can analyze the spherical constraint as 

\ba
T_i\,&=&\, a_i a_i - 1 \rightarrow \tilde{a}_i\,\tilde{a}_i\,-\,1 \nonumber \\
\,&=&\, [a_i\,-\,\partial_{a_i} (\theta_{kl}\,a_l \pi_k ]\,[a_i\,-\,\partial_{a_i} (\theta_{mn}\,a_n \pi_m ]\,-\,1 \nonumber \\
&=& a_i a_i \,-\,2a_i \partial_{a_i} (\theta_{kl} a_l \pi_k )\,-\,1
\ea

\noindent where we used that $\theta$ is infinitesimal.  This result motivate us to stress that the geometrical aspects of the Skyrme model are possibly affected by the NC parameter.  But a detailed study of the Skyrme geometry is not in the scope of this work.

It is important to notice that the commutative original version for the $SU(2)$ Skyrme Model can be restored. In fact, if we consider $\theta_{ij} \longrightarrow 0$, the Lagrangian (\ref{sk3}) is regained.

\subsection{The Noncommutative O(3) Nonlinear Sigma Model}

The O(3) nonlinear sigma model (NLSM) can be used as a theoretical laboratory that allows the investigation of some problems present in high energy physics, as the nonlinearity problems present in the quark and anti-quark interaction. Hence, we believe that it is relevant to analyze how the noncommutativity affects its original physical features.

In order to expand the perspective of this theoretical laboratory, we propose a NC version for the O(3) NLSM, which will be obtained via noncommutative Faddeev-Jackiw symplectic formalism (NCFJSF) \cite{sympnc}.

The dynamics of the O(3) NLSM is governed by the following
Lagrangian density
 
\be \label{00000} {\cal L}=\frac 12
(\partial_\mu\sigma)(\partial^\mu\sigma), 
\ee 

\noindent with the space-time metric $g_{00}=1$, $g_{ii}=-1$ and the O(3) spherical constraint
$\sigma_a\cdot\sigma_a=1$. 

To implement the noncommutative Faddeev-Jackiw symplectic formalism, the O(3) NLSM will be written in spherical
coordinates $(r,\theta,\phi)$, given by 

\be \label{000010}
\sigma=\left(\cos\theta,\sin\theta\cos\phi,\sin\theta\sin\phi\right),
\ee 

\noindent where the constraint was used, $r=1$. Thus, the Lagrangian density,
Eq. (\ref{00000}), written in spherical coordinates, is 

\be
\label{00020} {\cal L}=\frac 12\dot\theta^2
+\frac{\dot\phi^2\cdot\sin^2\theta}{2}-\frac{3}{2}, 
\ee 

\noindent which will be written in the phase-space coordinates $(\theta,\pi_\theta,\phi,\pi_\phi)$.  Namely, 

\be
\label{00030} {\cal L}^{(0)}=\pi_\theta\cdot
\dot\theta+\pi_\phi\cdot \dot\phi -\frac
12\pi_\theta^2-\frac{\pi_\phi^2}{2\cdot\sin^2\theta}-\frac{3}{2},
\ee 

\noindent where $\theta$ is different from $2n\pi$, with $n$ natural.

Following the prescription of the Faddeev-Jackiw symplectic formalism \cite{JBW}, the Dirac brackets
of the variables will be computed from the inverse of the symplectic matrix, 

\ba \label{00040} \left\{\theta(\vec
x),\pi_\theta(\vec y)\right\}&=&\left\{\phi(\vec x),\pi_\phi(\vec
y)\right\}=\delta^{(3)}(\vec x - \vec y),
\nonumber\\
\left\{\theta(\vec x),\theta(\vec y)\right\}&=&\left\{\phi(\vec x),\phi(\vec y)\right\}=\left\{\phi(\vec x),\theta(\vec y)\right\}=0,\nonumber\\
\left\{\pi_\theta(\vec x),\pi_\theta(\vec y)\right\}&=&\left\{\pi_\phi(\vec x),\pi_\phi(\vec y)\right\}=\left\{\pi_\phi(\vec x),\pi_\theta(\vec y)\right\}=0.\nonumber
\ea

After this point, the NC Faddeev-Jackiw symplectic formalism will be applied to achieve the NC version of the O(3) NLSM. 
Firstly, the inverse of symplectic matrix is proposed as being
\begin{widetext}
\be
\label{00050}
f^{-1}=\left(\begin{array}{llll}
\frac{\alpha}{2}\partial_{\vec{x}}\delta^{(3)}(\vec x-\vec z)& \quad\delta^{(3)}(\vec x - \vec z)&\qquad 0&\qquad\qquad 0\\
-\delta^{(3)}(\vec z-\vec x)& - \gamma\Theta(\vec x-\vec z)&\qquad 0&\qquad\qquad 0\\
\qquad 0&\qquad 0&\frac{\beta}{2}\partial_{\vec{x}}\delta^{(3)}(\vec x-\vec z)& \qquad\delta^{(3)}(\vec x - \vec z)\\
\qquad 0&\qquad 0&-\delta^{(3)}(\vec z-\vec x)&\qquad- \lambda\Theta(\vec x - \vec
z)\end{array}\right). \ee

Here, the notation $\partial_{\vec{x}}$
denotes $\prod_{i=1}^3\partial_i$.  After a straightforward
computation, the symplectic matrix is obtained as
\be \label{00060}
f=\left(\begin{array}{llll}
-\frac{2\gamma}{\Sigma}\Theta(\vec x - \vec y)&\quad\quad-\frac{2}{\Sigma}\delta^{(3)}(\vec x - \vec y)&\qquad\qquad 0&\qquad\qquad 0\\
\frac{2}{\Sigma}\delta^{(3)}(\vec y - \vec x)&\quad\quad\frac{\alpha}{\Sigma}\partial_{\vec{x}}\delta^{(3)}(\vec x - \vec y)&\qquad\qquad 0&\qquad\qquad 0\\
\qquad\qquad 0&\qquad\qquad 0&\quad-\frac{2\lambda}{\Gamma}\Theta(\vec x - \vec y)&\qquad-\frac{2}{\Gamma}\delta^{(3)}(\vec x - \vec y)\\
\qquad\qquad 0&\qquad\qquad 0&\quad\frac{2}{\Gamma}\delta^{(3)}(\vec x - \vec
y)&\qquad\frac{\beta}{\Gamma}\partial_{\vec{x}}\delta^{(3)}(\vec x - \vec
y)\\\end{array}\right).\ee 
\end{widetext}

\noindent Where $\Sigma = 2-8\alpha\lambda$ and
$\Gamma=2-8\beta\lambda$ are non-zero parameters; and $ \Theta(\vec
x) \equiv\prod_{i=1}^3\Theta(x_i)$, that is, the product of Signal functions.

After that, the one-form canonical
momenta will be computed. To this end, we will use the definition of
the symplectic matrix element, namely, 

\be \label{00070}
f_{\xi_i\xi_j}=\frac{\partial A_{\xi_i}}{\partial
\xi_j}-\frac{\partial A_{\xi_j}}{\partial \xi_i}. \ee 
Therefore, we have that
 \ba \label{00080}
A_\theta(\vec x)&=& \frac{2}{\Sigma}\pi_\theta(\vec x)+\frac{\gamma}{\Sigma}\int\Theta(\vec x - \vec z)\theta(\vec z) d^{3}\vec z,\nonumber\\
A_{\pi_\theta}(\vec x)&=& \frac{\alpha}{2\Sigma}\partial_x\pi_\theta(\vec x),\nonumber\\
A_\phi(\vec x)&=& \frac{2}{\Gamma}\pi_\phi(\vec x)+\frac{\lambda}{\Gamma}\int\Theta(\vec x - \vec z)\phi(\vec z) d^{3}\vec z,\\
A_{\pi_\phi}(\vec x)&=& \frac{\beta}{2\Gamma}\partial_x\pi_\phi(\vec
x).\nonumber \ea 

\noindent At this point, the new Lagrangian will be computed.
To this end, we assume that the canonical momenta are the one-form
momenta, {\it i.e.}, this implies that the model remains as 
second-order in velocity, then 

\be \label{00090}
A_{\pi_\theta}=A_{\pi_\phi}=0\,\,, 
\ee 

\noindent which imposes the following condition: $\alpha=\beta=0$. Hence, the canonical momenta are given by 

\ba \label{00100}
A_{\theta}&=&\pi_\theta+\frac{\gamma}{2}\int\Theta(\vec x-\vec
y)\theta(\vec y) d^{3}{\vec y},\\\nonumber
A_{\phi}&=&\pi_\phi+\frac{\lambda}{2}\int\Theta(\vec x-\vec
y)\phi(\vec y) d^{3}{\vec y}, 
\ea 

\noindent with the respective NC first-order
Lagrangian, 

\ba \label{00110} {\tilde {\cal L}} &=& \tilde
\pi_{\theta}\cdot \dot\theta+\tilde \pi_{\phi}\cdot \dot\phi
- \frac 12\pi_\theta^2- \frac {\pi_\phi^2}{2\cdot\sin^2\theta} -
\frac{3}{2}, 
\ea 

\noindent where $A_{\theta}={\tilde \pi}_\theta$ and
$A_{\phi}={\tilde \pi}_\phi$. Applying the transformations
$\pi_\theta\rightarrow{\tilde\pi}_\theta$ and
$\pi_\phi\rightarrow{\tilde\pi}_\phi$, and using (\ref{00100}), we have a
commutative first-order Lagrangian, namely,

\be 
\label{00120} 
{\tilde {\cal L}} = {\tilde \pi}_\theta\cdot
\dot\theta+\tilde \pi_\phi\cdot \dot\phi - \tilde V. 
\ee 

\noindent where $\tilde V$ is the NC symplectic potential, given by

\begin{widetext}
 \ba {\tilde V}
&=&\frac 12\tilde\pi_\theta^2+\frac {\tilde
\pi_\phi^2}{2\cdot\sin^2\theta}-\frac{\gamma}{2}\int\tilde\pi_\theta(\vec
x) \Theta(\vec x-\vec y)\theta(\vec y)d^{3} {\vec y} -
\frac{\lambda}{2\cdot\sin^2\theta}\int\tilde\pi_\phi(\vec
x)\Theta(\vec x-\vec y)\phi(\vec y)
d^{3}{\vec y} \\
\mbox{} &+&\frac{\gamma^2}{8}\left[\int\Theta(\vec x-\vec
y)\theta(\vec y)  d^{3}{\vec y}\right]^2
+\frac{\lambda^2}{8\cdot\sin^2 \theta}\left[\int\Theta(\vec x-\vec
y)\phi(\vec y) d^{3}{\vec y}\right]^2-\frac{3}{2}.\nonumber 
\ea

The Lagrangian (\ref{00120}) is commutative, however there are NC
contributions in the symplectic potential. In order to compute the
second-order Lagrangian in velocity, we compute the Euler-Lagrange
equation of motion for $\tilde\pi_\theta$ and eliminated this
variable in (\ref{00120}). Then we have the following second-order
Lagrangian
 
\be \label{00130} {\tilde {\cal L}}=
\frac{1}{2}\dot\theta^2+\frac{\dot\phi^2\cdot\sin^2\theta}{2}+\frac
{\gamma}{2}\int\dot\theta(\vec x)\Theta(\vec x - \vec y) \theta(\vec
y)d^{3}{\vec y}+\frac{\lambda}{2}\int\dot\phi(\vec x)\Theta(\vec x -
\vec y)\phi(\vec y)d^{3}{\vec y} - \frac{3}{2}. \ee 
\end{widetext}

Notice that in this case the noncommutativity introduces nonlocal terms which are functions of Signal functions. This result confirms the nonlocality imposed by the Moyal-Weyl product, differently from the anterior result (33) which is local.   Besides, we can see clearly that both angular coordinates are affected as seen in (47).  A target for future investigation would be  if the noncommutativity would modify the action in (34) with $r\not= 1$.  Namely, if the action (36) had a $r$-term, how this term would be modified by the NC procedure?    Looking at (47), we can ask ourselves how the geometry of NLSM would be modified.
The original commutative model is restored if the NC parameters are zero.

\section{conclusion}

In this Letter we obtained Lagrangian formulations for the NC versions of the reduced $SU(2)$ Skyrme model and the $O(3)$ nonlinear sigma model, which are two important nonlinear systems. 
The first one originally describes baryons and their interactions.  And the second one describes quarks and anti-quarks interactions.  Both are related by the soliton solution of NLSM.  Here we introduced the representations of both models at Planck's scale.

This was achieved through an alternative new way to obtain NC models based on the Faddeev-Jackiw symplectic formalism. An interesting feature about this formalism lies on the symplectic structure, which is defined at the beginning of the process. The choice of the symplectic structure defines the NC geometry of the model and the Planck´s constant enters the theory via Moyal-Weyl product. 

This formalism also describes precisely how to obtain a Lagrangian 
description of the NC version of the system.  As we know, NC field 
theories provide fruitful avenues of exploration for several reasons \cite{szabo, jackiw}. 
Firstly, some quantum field theories have a better behavior in noncommutative spacetime 
than in ordinary spacetime. In fact, some are completely finite, even non-perturbatively. 
Thus, spacetime noncommutativity presents itself as an alternative to string theory or 
supersymmetry. 

Secondly, it is a useful arena for studying physics beyond the standard 
model and also for standard physics in strong external fields. Thirdly, it sheds light 
on alternative underlying issues in quantum field theory. For instance, renormalization 
and axiomatic programs. And finally, it naturally relates field theory to gravity. Since the field theory can be quantized, 
this may provide significant insights into the problem of quantizing gravity. 

Another investigation can reveal if there is a relation between the soliton solution of the NC NLSM and the baryons described by the NC Skyrme model.  We do not know if the noncommutativity preserves the commutative relation as $\theta \rightarrow 0$.

\section{Acknowledgments}
 
\noindent EMCA would like to thank the hospitality and kindness of Dept. of Physics 
of Federal University of Juiz de Fora where part of this work was done.
The authors would like  to thank CNPq and FAPEMIG (Brazilian research agencies) 
for financial support.


\begin{thebibliography} {99}

\bibitem{snyder}  H. S. Snyder, Phys. Rev. 71 (1947) 38.

\bibitem{yang}  C. N. Yang, Phys. Rev. 72 (1947) 874.


\bibitem{strings} A. Connes, M. Douglas and A. Schwarz, JHEP 02 (1998) 003. For recent reviews: 
N. A. Nekrasov, hep-th/0011095; A. Konechny and A. Schwarz, hep-th/0012145; J. A. Harvey, hep-th/0102076; N. Seiberg and E. Witten, JHEP 09 (1999) 032.

\bibitem{banerjee}   R. Banerjee, Mod. Phys. Lett. A 17 (2002) 631.

\bibitem{thooft}   G. 't Hooft, Class. Quant. Grav. 16 (1999) 3263; hep-th/0003005; hep-th/0105105.

\bibitem{string} E. M. C. Abreu, A. A. Deriglazov, C. Filgueiras, C. Neves, W. Oliveira and C. Wotzasek, Phys. Rev. D 76 (2007), 064007.

\bibitem{witten} N. Seiberg and E. Witten, JHEP 09 (1999) 032.

\bibitem{other}   A. A. Deriglazov, Phys. Lett. B 530 (2002) 235; Phys. Lett. B 555 (2003) 83.

\bibitem{alexei}  A. A. Deriglazov, Noncommutative version of an arbitrary nondegenerated mechanics, hep-th/0208072.

\bibitem{belov}  I. Y. Arefeva, D. M. Belov, A. A. Giryavets, A. S. Koshelev, P. B. Medvedev, Noncommutative field theories and (super) string field theories, hep-th/0111208.

\bibitem{szabo} R. J. Szabo, Quantum gravity, field theory and signatures of noncommutative spacetime, arXiv:09062913; Quantum field theory on noncommutative spaces,  hep-th/0109162.

\bibitem{omer}   O. F. Dayi, L. T. Kelleyane, Mod. Phys. Lett. A 17, 1937 (2002).

\bibitem{RB}  R. Amorim and J. Barcelos-Neto, Phys. Rev. D 64 (2001) 065013, and references therein.
  
\bibitem{hp}  P. A. Horv\'athy and M. S. Plyushchay, Phys. Lett. B 595 (2004) 547; JHEP 06 (2002) 033.

\bibitem{dh}  C. Duval and P. A. Horv\'athy, Phys. Lett. B 479 (2000) 284.

\bibitem{peierls}   R. Peierls, Z. Phys. 80 (1933) 763; D. Hofstader, Phys. Rev. B 14 (1976) 2239.

\bibitem{dh2}  C. Duval and P. A. Horv\'athy, J. Phys. A 34 (2001) 10097; P. A. Horv\'athy, Ann. Phys. 299 (2002) 128.

\bibitem{djt}   G. V. Dunne, R. Jackiw and C. A. Trugenberger, Phys. Rev. D 41 (1990) 661.

\bibitem{mrs}  S. Minwalla, M. Van Raamsdonk and N. Seiberg, JHEP 02 (2000) 020. 

\bibitem{DJEMAI1}   A. E. F. Djema\"\i $\;$ and H. Smail, Commun. Theor. Phys. 41 (2004) 837; On noncommtative classical mechanics, hep-th/0309034.

\bibitem{DJEMAI2}   A. E. F. Djema\"\i, Int. J. Theor. Phys. 35 (1996) 519.

\bibitem{amorim}   R. Amorim, Phys. Rev. Lett. 101 (2008) 081602.

\bibitem{sigma}  E. M. C. Abreu, A. C. R. Mendes and W. Oliveira, SIGMA 6 (2010) 083.


\bibitem{sympnc} E. M. C. Abreu, C. Neves and W. Oliveira, Int. J. Mod. Phys. A 21 (2006) 5359-5369.

\bibitem{amo}  For a review of the method, see E.M.C. Abreu, A.C.R. Mendes, W. Oliveira, SIGMA 6 (2010) 059.

\bibitem{JBW}  L. Faddeev and R. Jackiw, Phys. Rev. Lett. 60 (1988) 1692.









\bibitem{MOYAL1}   J.E. Moyal, Proc. Camb. Phil. Soc. 45 (1949) 99; J. Vey, 
Commentari Mathematici Helvetici 50 (1975) 421; M. Flato {\it et al}, 
Commentari Mathematici Helvetici 31 (1975) 47; M. Flato {\it et al}, J. Math. Phys. 17 (1976) 1754; F. Bayen {\it et al}, Ann. of Phys. 111 (1978) 61.

\bibitem{CNWO}  C. Neves and W. Oliveira, Phys. Lett. A 321 (2004) 267.

\bibitem{sky1} G. Adkins, Chiral Solitons, ed. Keh-Fei Liu (World Scientific, Singapore, 1987); G. S. Adkins, C. R. Nappi and E. Witten, Nucl. Phys. B 228 (1983) 552.

\bibitem{sky2}  J. Ananias Neto, C. Neves and W. Oliveira, Phys. Rev. D 63, 085018.

\bibitem{bazeia}   D. Bazeia, Mod. Phys. Lett. A 6 (1991) 1147.

\bibitem{FJ}   R. Floreanini and R. Jackiw, Phys. Rev. Lett. 59 (1987) 1873.

\bibitem{jackiw}   Z. Guralnik, R. Jackiw, S.Y. Pi, A.P. Polychronakos, Phys. Lett. B 517 (2001) 450, hep-th/0106044.



\end{thebibliography}
\end{document}